\documentstyle[seceq,epsf,supplement]{ptptex}

\newcommand{\beq}{\begin{eqnarray}}
\newcommand{\eeq}{\end{eqnarray}}
\newcommand{\sbeq}{\begin{subeqnarray}}
\newcommand{\seeq}{\end{subeqnarray}}
\newcommand{\do}{^{\circ }}
\newcommand{\eps}{\epsilon}

\newcommand{\btem}{\bibitem}
\newcommand{\YO}{Y. Oono}
\newcommand{\NG}{N. Goldenfeld}

\title{
Renormalization-group Resummation of a Divergent Series of 
the Perturbative Wave Functions of Quantum Systems
}

\author{
Teiji {\sc Kunihiro}\footnote{E-mail address: kunihiro@rins.ryukoku.ac.jp}
}

\inst{
Faculty of Science and Technology, Ryukoku University,
Otsu, 520-2194 Japan
}

\recdate{
\today
}

\abst{
The perturbative renormalization group(RG) equation
 is applied to resum divergent series of perturbative wave functions
 of quantum anharmonic oscillator.
 It is found that the resummed series 
 gives  the {\em cumulant}  of the naive perturbation series.
It is shown that a reorganization of the resummed series 
reproduce the correct asymptotic
 form of the wave function at $x\rightarrow \infty$
 when the perturbation expansion is stopped at the fourth order.
A brief comment is given on the relation between 
 the present method and the delta-expansion method, which is based on 
 a kind of a nonperturbative RG equation.
}

\begin{document}
\newcommand\theHequation {\theHsection.\arabic{equation}}

\maketitle

\section{Introduction}
\setcounter{equation}{0}
In this talk, we shall present a novel use
 of the perturbative renormalization
 group equations  $\grave{a}$ la Gell-Mann-Low \cite{rg} to solve problems
 in  other fields than quantum field theory and statistical mechanics.

When applying a perturbation theory to any problem,  one 
 will encounter  divergent or at best asymptotic series.
 One needs to resum the divergent series to obtain a
 sensible  result from the perturbation theory.
 Thus various resummation techniques  have been devised.\cite{bo}\ 
Recently, 
a unified and mechanical method for global and asymptotic analysis
 has been  proposed by Goldenfeld et al.\cite{cgo}\  This is called the
 renormalization group (RG) method. Indeed the RG equations 
$\grave{a}$ la Gell-Mann-Low is known to have a peculiar power to improve
 the global nature of 
functions obtained in the perturbation theory in quantum field 
theory.\cite{rg}\ 
A unique feature  of Goldenfeld et al's method is to start with the naive 
 perturbative expansion and allow  secular terms to appear in contrast 
with all previous methods;\cite{bo}\ adding unperturbed solutions to the 
 perturbed solutions so that the secular terms
 vanish at a ``renormalization point" $t=t_0$ and then  applying the
 RG equation, one obtains a resummed perturbation series.

Subsequently, the RG method has been formulated on the basis
 of the classical theory of envelopes;\cite{kuni1,kuni2}\ 
 it has been indicated \cite{kuni1} that the notion of 
envelopes is also  useful for improving the effective potential 
 in quantum field theory.\cite{effective}
\footnote{
Shirkov has extracted the notion of {\em functional self-similarity} 
 (FSS) as the essence of the RG.\cite{shirkov}\ 
We remark  that the notion of FSS is only applicable to autonomous
 equations, while that of envelopes 
is equally applicable to non-autonomous equations, too.}

 The purpose of the present work is to make an introductory account
 of the RG method as formulated in Ref.'s~\citen{kuni1,kuni2}
 and  apply the method  to Schr\"{o}dinger equation of the quantum 
anharmonic oscillator (AHO) to  resum the naive perturbation series by 
the RG equation; we shall also examine whether our method is useful to
 obtain the  asymptotic form of the wave functions.\cite{rapid}
 
AHO has been a theoretical laboratory for examining the validity of various 
approximation techniques.\cite{rev1}
 Bender and Bettencourt \cite{bb} have recently shown that  multiple-scale
 perturbation theory (MSPT) can be successfully 
 applied to the quantum anharmonic oscillator;
 see also Ref.~\citen{frasca1}. 
Using the fourth order perturbation series,
they  obtained  an asymptotic
 behavior of the wave function $\psi (x)$ for large $x$ which 
 is in good  agreement with  the WKB result. 
One should remark  here that  a further resummation had to be  adopted 
 for the latter case, which is not intrinsic in 
MSPT and a similar method had been proposed by Ginsburg and Montroll
\cite{gm}. 

We shall show that the resummation of the  perturbation series of the  
wave functions is performed in the RG method more
 mechanically and explicitly than in MSPT:
\footnote{
As for the application of the RG method a l\`{a} Goldenfeld et al,
to the  time-dependent Schr\"{o}dinger equations, 
see Ref.~\citen{egus,frasca2}.}
Since our method is easy to perform to higher orders of the perturbative
 expansion, it is examined how the results of Bender and Bettencourt 
persist or are modified at the higher orders.

In the next section, an  introduction is given of the 
RG method for improving perturbative solutions of differential equations.
 Then the method is applied to  quantum mechanics in \S 3.
 The final section is devoted to a summary and remarks.

\setcounter{equation}{0} 
\section{The RG method for global analysis;  simple examples}
\renewcommand{\theequation}{\thesection.\arabic{equation}}

In this section, using  simple examples, we shall show how the RG 
method\cite{cgo} works for obtaining  global and asymptotic behavior of 
solutions of differential equations. 
One will see that the reasoning for various steps in the procedure
 and the underlying picture are quite different from the
 original ones given in Ref.~\citen{cgo}: The  present
 formulation  emphasizes the role of initial conditions and the relevance
 to envelopes of perturbative local solutions.\cite{kuni1,kuni2}

\subsection{A linear equation}

Let us first take the following simplest example,\cite{kuni1}
\beq
\frac{d^2 x}{dt^2}\ +\ \eps \frac{dx}{dt}\ +\ x\ =\ 0,
\label{eq:2.1}
\eeq
where $\eps$ is supposed to be small. The solution to (\ref{eq:2.1}) reads
$ x(t)= \bar {A} \exp (-\eps t/2)$
 $\cdot\sin( \sqrt{1-\eps^2/4}t+ \bar {\theta}),$
where $\bar {A}$ and $\bar{\theta}$ are constants.

 Now, let us  obtain the solution around the initial time $t=t_0$ in a
 perturbative way,  expanding $x$ as
\beq
x(t, t_0) = x_0(t, t_0) \ +\ \eps  x_1(t ,t_0)\ +\ \eps ^2 x_2(t, t_0)\ 
+\ ... .\label{eq:2.2}
\eeq
The initial condition may be specified by
\beq
x(t_0, t_0)= W(t_0).\label{eq:2.3}
\eeq
We suppose that the initial value $W(t_0)$ is always on an exact solution
 of Eq.(2.1) for any  $t_0$. We also expand the initial value $W(t_0)$;
\beq
W(t_0) = W_0(t_0) \ +\ \eps  W_1(t_0)\ +\ \eps ^2 W_2(t_0)\ 
+\ ... ,
\eeq
and the terms $W_i(t_0)$ will be determined so that the perturbative solutions
 around different initial times $t_0$  have an 
 envelope.  Hence the initial value $W(t)$ thus constructed will give 
the (approximate but) global solution of the equation. 

Let us perform the above program. 
The lowest solution may be given by
\beq
x_0(t, t_0) = A(t_0)\sin (t +\theta (t_0)), 
\eeq
where it has been made it explicit that the constants $A$ and $\theta$ may 
depend on the initial time $t_0$.
The initial value $W(t_0)$ as a function of $t_0$
 is specified as
\beq
W_0(t_0)= x_0(t_0,t_0)= A(t_0)\sin (t_0 +\theta (t_0)).
\eeq
We remark that Eq.(2.5) is a neutrally stable solution; with 
 the perturbation $\eps \not=0$ 
 the constants $A$ and $\theta$ may move slowly.  We shall see that 
 the RG equation, which will be interpreted as the envelope equation,
\cite{kuni1}\ gives the equations describing the slow motion of
 $A$ and $\theta$.
 
The first order equation now reads
$\ddot{x}_1\ +\ x_1\ =- A\cos(t+\theta),$
 and we choose the solution in the following form,
\beq
x_1(t, t_0)= -{A}/{2}\cdot (t -t_0)\sin(t+\theta),
\eeq 
which means that the first order initial value $W_1(t_0)=0$ so that
the lowest order value $W_0(t_0)$ approximates the exact value as closely 
 as possible.
Similarly, the second order solution may be given by
\beq
 x_2(t)= 
{A}/{8}\cdot \{ (t-t_0)^2\sin(t +\theta) - (t-t_0)\cos(t+\theta)\},
\eeq
 thus $W_2(t_0)=0$ again for the present linear equation.

It should be noted  that  the secular terms have appeared 
 in the higher order terms, which are  absent in the 
exact solution and invalidates the perturbation theory for $t$ far
  from $t_0$. However, with the very existence of the secular terms,
 we could make $W_i(t_0)$ ($i=1, 2$) vanish and  $W(t_0)=W_0(t_0)$
 up to the third order.

Collecting the terms, we have 
\beq
x(t, t_0)&=& A\sin (t +\theta) -\eps{A}/{2}\cdot (t -t_0)\sin(t+\theta)
  \nonumber \\ 
 \ \ \ & \ \ \ & +\eps^2{A}/{8}\cdot 
\{ (t-t_0)^2\sin(t +\theta) - (t-t_0)\cos(t+\theta)\},
\eeq
and more importantly
\beq
W(t_0)=W_0(t_0)=A(t_0)\sin (t_0 +\theta (t_0))
\eeq
 up to $O(\eps^3)$. Notice that $W(t_0)$ describing the solution 
 is parametrized by possibly slowly moving variable $A(t_0)$ and 
 $\phi (t_0)\equiv t_0+\theta (t_0)$ in a definite way.

Now we have a family of curves $\{C_{t_0}\}_{t_0}$ given by functions 
$\{x(t, t_0)\}_{t_0}$ parametrized with $t_0$. 
 They are all on the exact curve $W(t)$ at $t=t_0$ 
 up to $O(\eps ^3)$, but  only valid locally for $t$ near $t_0$. 
 So it is conceivable that the envelope  $E$ of $\{C_{t_0}\}_{t_0}$ which 
 contacts with each local solution at $t=t_0$ will give a global solution.
 Thus the envelope function $x_{_E}(t)$ coincides with $W(t)$;  
\beq
x_{_E}(t)=x(t,t)=W(t).
\eeq  
Our task is actually to determine $A(t_0)$ and $\theta(t_0)$ as 
 functions of $t_0$ so that the family of the local solutions has an 
 envelope. According to the classical theory of envelopes, 
the above program can be achieved by imposing that the envelope equation 
\beq
 \frac{dx(t, t_0)}{d t_0}=0,
\eeq
 gives the solution $t_0=t$.\cite{kuni1}\ 
Inserting (2.9) into (2.12), we have
\beq
\frac{dA}{dt_0} + \eps A =0,  \ \ \ 
\frac{d\theta}{dt_0}+\frac{\eps^2}{8}=0,
\eeq
where we have utilized the fact that $dA/dt$ is $O(\eps)$ and neglected
 the terms of $O(\eps^3)$.
Solving the equations, we have 
\beq
A(t_0)= \bar{A}{\rm e}^{-\eps t_0/2}, \ \ \ 
\theta (t_0)= -{\eps^2}/{8}\cdot t_0 + \bar{\theta},
\eeq
where $\bar{A}$ and $\bar{\theta}$ are constant numbers.
 Thus we obtain 
\beq
x_{_E}(t)= x(t, t)=W_0(t)= 
\bar{A}\exp(-{\eps}/{2}\cdot t)\sin((1-{\eps ^2}/{8})t + 
\bar{\theta}),
\eeq
up to $O(\eps^3)$.
Noting that $\sqrt{1 - {\eps^2}/{4}}= 1 - {\eps^2}/{8} + O(\eps ^4)$, 
 one finds
 that the resultant envelope function $x_{_E}(t)=W_0(t)$ is an approximate but 
{\em  global} solution to Eq.(2.1).

\subsection{Forced Duffing Equation}
As an example of nonlinear equations, we take the forced Duffing 
equation where the frequency of the external force is near the intrinsic 
one;\cite{kuni2}\footnote{
There are some errors in the Appendix of Ref.~\citen{kuni2},
 which are corrected below.}
\sbeq
\ddot {x}+ 2\eps \gamma \dot{x}+ (1+\eps \sigma)x + \eps hx^3&=&
\eps f\cos t,  \\
\ddot {y}+ 2\eps \gamma \dot{y}+ (1+\eps \sigma)y + \eps hy^3 &=&
\eps f\sin t.\label{sduffing}
\seeq
When $\gamma =f=0$, the equation describe the usual (classical) 
anharmonic oscillator with the quartic potential.
Defining a complex variable $z=x+i y$, one has
\beq
\ddot {z}+ 2\eps \gamma \dot{z}+ (1+\eps \sigma)z +
{\eps h}/{2}\cdot 
 (3\vert z\vert^2z +{z^{\ast}}^3)= \eps fe^{it}.\label{duffing}
\eeq
We suppose that $\eps$ is small.

Expanding $z$ as
$z(t, t_0)=z_0(t, t_0)+\eps z_1(t, t_0) +\eps ^2z_2(t, t_0) + \cdots$, 
we first, as before,   obtain a perturbative solution around the initial 
time $t=t_0$, with the initial condition
\beq
z(t_0, t_0)=W(t_0).
\eeq
The initial value $W(t_0)$ is to be determined so as to the
 perturbative solutions at $t_0$ varied continue smoothly.
We thus also expand the initial value as
$W(t_0)=W_0(t_0)+\eps W_1(t_0) +\eps ^2W_2(t_0) + \cdots$.

A simple manipulation gives the  solution up in the first order
 approximation as
\beq
z(t; t_0)&=& 
{\cal A}(t_0)e ^{it}-i{\eps}/{2}\cdot (t-t_0)
[f- {\cal A}(t_0)(\sigma + 2i\gamma)  \nonumber \\ 
 \ \ & & -{3h}/{2}\cdot \vert {\cal A}(t_0)\vert^2{\cal A}(t_0)]e ^{it}
 + \eps /{32}\cdot {{\cal A}^{\ast}(t_0)}^3e^{-3it} + O(\eps^2).
\eeq
This solution implies that we have chosen the initial condition as
\beq
W(t_0)=z(t_0,t_0)=
{\cal A}(t_0)e ^{it}+
 \eps /{32}\cdot {{\cal A}^{\ast}(t_0)}^3e^{3it} + O(\eps^2).
\eeq
The RG or envelope  equation reads
\beq
\frac{d z}{d t_0}=0, 
\eeq
with $t_0=t$, which leads to
\beq
\dot{{\cal A}}=i\eps/{2}\cdot 
[(\sigma +2i\gamma){\cal A} +
{3h}/{4}\cdot\vert {\cal A}\vert ^2{\cal A}-f],  
\eeq
 up to $O(\eps ^2)$.  Here we have discarded terms such as $\eps d{\cal A}/dt$,
 which is $O(\eps ^2)$ because $d{\cal A}/dt=O(\eps)$.
 The envelope is given
\beq
z_E(t)=z(t; t_0=t)=
 {\cal A}(t)e ^{it} + \frac{\eps}{32} {{\cal A}^{\ast}}^3e^{-3it} + O(\eps^2).
\eeq
We identify $z_E(t)$ with a global solution of Eq.(\ref{duffing}); 
$x(t)={\rm Re}[z_E]$ and $y(t)={\rm Im}[z_E]$ are solutions to 
 Eq.(\ref{sduffing}).
If we parametrize the amplitude as 
${\cal A}(t)=R(t){\rm exp}(i\theta (t))$, we have
\sbeq
x(t)&=&R(t)\cos (t +\theta (t))+ \eps\frac{h}{32}R(t)^3\cos 3(t+\theta(t)),
  \\  
y(t)&=&R(t)\sin (t +\theta (t))- \eps\frac{h}{32}R(t)^3\sin 3(t+\theta(t)),
\label{ssol}
\seeq
where the real functions $R(t)$ and $\theta (T)$ satisfy the 
 following coupled equation;
\sbeq
\dot{R}(t)&=&-\frac{\eps}{2}[\sqrt{\sigma ^2+4 \gamma ^2}\sin \alpha R(t)
+f\sin \theta(t)], \\
R(t)\dot{\theta}(t)&=& \frac{\eps}{2}[\sqrt{\sigma ^2+4 \gamma ^2}
\cos \alpha R(t)+ \frac{3}{4}hR(t)^3-f\cos \theta (t)],\label{couple}
\seeq
 with $\tan \alpha =2\gamma/\sigma$.

 When the external force is absent, i.e., $f=0$, Eq.(\ref{couple})
 is readily solved, yielding
\beq
R(t)=R_0e^{-\eps\tilde{\gamma}t}, \ \ \ 
\theta (t)= \Delta \omega t- \frac{3h}{8\tilde{\gamma}}R_0^2 
\{e^{-2\eps\tilde{\gamma}t}-1\} +\theta_0,
\eeq
with $\tilde{\gamma}=1/2\cdot\sqrt{\sigma ^2 +4\gamma ^2}\sin \alpha$, 
$\Delta \omega =\eps\sqrt{\sigma ^2+4\gamma^2}\cos \alpha$,
 $R_0$ and $\theta_0$ being constant.\footnote{
When $\gamma =0$, $R(t)=R_0={\rm const.}$ and 
$\theta(t)=\eps (\sigma+ 3hR_0^2/4)t +\theta_0$.
}\ 
Inserting $R(t)$ and $\theta(t)$ into $x(t)$ and $y(t)$ in 
(\ref{ssol}), one has an explicit form of the solution to the anharmonic 
oscillator.
Notice that the solution contains an infinite order of $\eps$,
 although we have only made a first order calculation.
When $f\not=0$, the coupled equation can not be solved by quadrature
  hence the equation is not reduced to a simpler equation than the
 original ones, which may be a reflection that the forced Duffing equation
 gives chaotic solutions.
 
Extensive applications of the RG method to nonlinear equations including 
 partial differential equations and difference equations
  are given in Ref.'s~\citen{cgo,kuni1,appl}.

\section{The RG-resummation of divergent perturbative wave functions}

Now let us apply the RG method to a quantum mechanical problem.
We take  the anharmonic oscillator\footnote{
We use the notations of Ref. 10) for later comparison.}
\beq
(H- E)\psi(x)=0, \ \ 
H=p^2 + x^2/4 +\eps x^4/4,
\eeq
with the boundary condition
 $\psi(\pm \infty)=0$.
WKB analysis shows that for large $x$, 
\beq
\psi(x)\sim {\rm exp}\{-\sqrt{\eps}\vert x\vert ^3/6\}.\label{wkb}
\eeq
We are interested in  how the perturbation theory
 can reproduce the WKB behavior or not.
  
\newcommand{\ee}{e^{-x^2/4}}
\subsection{Preliminaries}
We  first  apply the Bender-Wu method \cite{bw} for 
 performing Rayleigh-Schr\"{o}dinger perturbative expansion:
\beq
\psi(x) \sim \sum_{n=0}^{\infty}\eps^n y_n(x)\ \ \ {\rm and}
\ \ \ E(\eps) \sim \sum_{n=0}^{\infty}\eps^n E_n.
\eeq
The boundary condition is taken as
$y_0(0)=1\ \ \ {\rm and}\ \ \ y_{n\geq 1}(0)=0.$
The lowest order solution reads 
 $y_0(x)=e^{-x^2/4}\ \ \ {\rm and} \ \ \ E_0=1/2.$
The higher order terms with $n\geq 1$ are written as
\beq
 y_n(x)=\ee P_n(x),\label{secular}
\eeq
where $P_n(x)$ is a polynomial, which 
satisfies the recursion relation;
\beq
P_n''(x) -x P_n'(x)=\frac{x^4}{4}P_{n-1}(x) - \sum _{j=0}^{n-1}P_j(x)E_{n-j}.
\label{recursion}
\eeq
We remark  that
 $y_n(x)$ in Eq.(\ref{secular}) may be identified as a secular term because
 it is a product of the unperturbed solution and a function that
  increases as $x$ goes large.

The eigenvalues $E_n$ are given by
\beq
E_n=\frac{1}{2\sqrt{\pi}}\int_{-\infty}^{\infty}dx \ y_0(x)
\{
\frac{x^4}{4}P_{n-1}(x) - \sum _{j=1}^{n-1}P_j(x)E_{n-j}
\}.\label{eigen}
\eeq
Notice that $E_n$ is determined in terms  of  only the polynomials
 $P_j(x)$ with $j\leq n-1$. 
  
With these eigenvalues $E_j$ \ ($j\leq n$), 
 $P_n(x)$  is determined by Eq.(\ref{recursion}).
 The polynomials $P_n(x)$ up to the six order are  
 presented in Ref.10), which we refer to.

\subsection{The RG method}
Now we apply the renormalization group method as formulated in \S 2.

We first calculate the wave function $\psi(x;x_0)$
around an initial point $x=x_0$ in a perturbative way:
\beq
\psi(x;x_0)\sim 
\sum_{n=0}^{\infty}\eps^n \varphi_n(x; x_0)\ \ \ {\rm and}
\ \ \ E(\eps) \sim \sum_{n=0}^{\infty}\eps^n E_n,
\eeq
with  the initial or boundary condition (BC) at $x=x_0$;
\beq
\psi(x_0; x_0)= W(x_0).
\eeq
We suppose that the boundary value $W(x_0)$ is always on an exact 
solution of Eq.(3.1).
We also assume that $W(x_0)$ can  be also expanded in a power series of $\eps$
\beq
W(x_0)=\sum_{n=0}^{\infty}\eps^n W_n(x_0)
\eeq

If we stop at a finite order, say, the $N$-th order, we will have
$\sum_{n=0}^{N}\eps^n \varphi_n(x)\equiv \psi^{(N)}(x; x_0)$
 which is valid only locally at $x\sim x_0$.
However, one may take the following geometrical
 point of view; namely,
 we have now a family of curves $\{\psi^{(N)}(x; x_0)\}_{x_0}$
 parametrized with $x_0$, and each curve of the family is a good approximation
 around $x=x_0$. Then, if each curve is continued smoothly, the
 resultant curve will valid in a global domain of $x$.
 This is nothing else than to construct the envelope of the family of
 curves.
 More specifically, we only have to determine 
the boundary values $W(x_0)$  so that the perturbative solutions 
around $x=x_0$ form an envelope; see Fig.1.
Furthermore, to be as accurate as possible, the lowest value 
$W_0(x_0)$ should approximate the exact value $\psi(x_0, x_0)$ as close
 as possible, or $W_{n\geq 1}(x_0)$ should be made as small as possible.

\begin{figure}
\epsfxsize= 6.5 cm
\centerline{\epsfbox{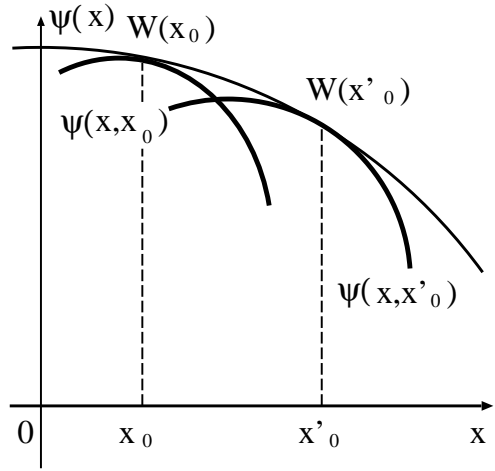}}
\caption{The local solutions and the global solution as their envelope.}
\label{fig1}
\end{figure}

The above program can be performed as follows.
The decisive observation is that the lowest order solution may be written as
\beq
\varphi_0(x; x_0)=Z(x_0)e^{-x^2/4},
\eeq
with an arbitrary function $Z(x_0)$ of $x_0$.\footnote{
It may be noted that the amplitude $Z(x_0)$ and the boundary position 
$x_0$ correspond
 to the  wave function renormalization constant and the renormalization
 point in the field theory, respectively.}
We remark that the
$E_0=1/2$ in this case, too.
Accordingly, the boundary value is
\beq
W_0(x_0)=Z(x_0)e^{-x_0^2/4}.
\eeq

The higher order terms may be  written as
\beq
\varphi_n(x; x_0)=Z(x_0)\ee Q_n(x; x_0),
\eeq
where $Q_n(x; x_0)$ is a polynomial of $x$, dependent on $x_0$.
$Q_n(x; x_0)$ is found to 
satisfy the same recursion relation Eq.(\ref{recursion}) as $P_n(x)$.
We impose the boundary condition (B.C.) as
\beq
\varphi_{n}(x_0; x_0)=W_n(x_0)=0 \ \ \ {\rm or}\ \ \  
Q_{n}(x_0; x_0)=0,\label{bc}
\eeq
 for $n\geq 1$ because we want to make the lowest order solution 
 as close as the exact one at $x=x_0$.\footnote{
We remark that the B.C.(\ref{bc}) or Eq.(\ref{bc2}) below corresponds to
 a renormalization condition in the field theory.}
\newcommand{\calP}{Q}
Thus one readily obtains 
\beq
\calP _1(x; x_0)= P_1(x) -P_1(x_0),\label{first}.
\eeq
We remark that a constant is the solution of the homogeneous equation.\footnote{
We remark that the second term of Eq.(\ref{first}) corresponds to a counter
 term in the renormalization procedure in the field theory.}
The second order equation reads
\beq
\calP_2''(x;x_0) -x \calP_2'(x;x_0)&=& 
 \bigl(\frac{x^4}{4}P_1(x)-\sum_{j=0}^{2}P_{j}(x)E_{n-j}\bigl)\nonumber \\  
 \ \ & & -P_1(x_0)({x^4}/{4}-E_1).\label{second} 
\eeq
One can verify that $E_2$ is given by Eq.(\ref{eigen}).
Since Eq.(\ref{second}) is a linear equation  and the inhomogeneous part
is a linear combination of those for $P_2(x)$ and 
 $P_1(x)$,  $\calP_2(x; x_0)$ satisfying (\ref{bc}) is  given by 
\beq
\calP _2(x; x_0)= (P_2(x) - P_2(x_0))-P_1(x_0)(P_1(x) -P_1(x_0)).
\eeq
Repeating the procedure,
one finds that $Q_n(x; x_0)$ are expressed in terms of 
$P_j(x)$ ($j\leq n$); see Ref.~\citen{rapid} for the explicit forms
 for $n\leq 6$.

Thus we obtain the approximate solution valid around $x\sim x_0$,
\beq
\psi (x; x_0)\sim Z(x_0)\ee \sum _{n=0}^{\infty}\eps^n \calP _n(x; x_0),
\eeq
where
\beq
\psi(x_0; x_0)=W_0(x_0)=Z(x_0)e^{-x_0^2/4}.\label{bc2}
\eeq

Now we have obtained a family of 
 curves $\{\psi(x;x_0)\}_{x_0}$ with $x_0$ parametrizing the curves;
 each curve is a good approximation for $x$ around $x_0$.
The envelope of the local solutions is giving by 
 solving the following equation,
\beq
\frac{d\psi (x; x_0)}{dx_0}\Bigl\vert _{x_0=x}=0,
\eeq
which is in the same form as
 the renormalization group equation; one may have called it the RG equation.
The equation gives a condition which $Z(x_0)$ must satisfy;
\beq
\frac{dZ}{dx}= Z(x)\frac{d}{dx_0}\sum _{n=0}^{\infty}
\eps^n(- \calP _n(x; x_0))
\Bigl \vert _{x_0=x}.
\eeq
Defining $f_n(x)$ by 
\beq
-\frac{d}{dx_0}\calP _n(x; x_0)\bigl \vert _{x_0=x}= 
\frac{df_n(x)}{dx},
\eeq 
one obtains 
\beq
Z(x)= {A}\cdot {\rm exp}[\sum _{n=0}^{\infty}\eps ^n f_n(x)],
\eeq 
where $A$ is a constant. 
With this solution, the global solution  $\psi _{E}(x)$ is given 
by the boundary value by construction;
\beq
\psi _{E}(x)= W_0(x)&=& Z(x)\ee =
 {A}\cdot {\rm exp}[-\frac{x^2}{4} +\sum _{n=1}^{\infty}\eps ^n f_n(x)] .
\eeq
This is one of the main results of the present work.

$f_n(x)$'s are easily calculated in terms of $P_n(x)$;
\beq
f_1(x)&=& P_1(x)=-\frac{3}{8}x^2-\frac{1}{16}x^4, 
f_2(x)= P_2(x)-\frac{1}{2}P_1(x)^2=  \frac{21}{16}x^2+\frac{11}{64}x^4
 +\frac{1}{96}x^6,\nonumber   \\ 
f_3(x)&=&P_3(x)-P_1(x)P_2(x)+\frac{1}{3}P_1(x)^3 = 
- \frac{333}{32}x^2-\frac{45}{32}x^4-
  \frac{21}{192}x^6-\frac{1}{256}x^8,\nonumber \\ 
f_4(x)&=&P_4(x)-P_1(x)P_2(x)-\frac{1}{2}P_2(x)^2+P_1(x)^2P_2(x) -
\frac{1}{4}P_1(x)^4,\nonumber \\ 
\ \ &=&  \frac{30885}{256}x^2 + \frac{8669}{512}x^4 + \frac{1159}{768}x^6  
       + \frac{163}{2048}x^8 + \frac{x^{10}}{512},\nonumber \\ 
f_5(x)&=&P_5(x) -P_1(x) P_4(x) - P_2(x) P_3(x)+P_1(x) P_2(x)^2 
        -P_1(x)^3 P_2(x)\nonumber \\
 \ \ & &  + P_1(x)^2 P_3(x) + \frac{1}{5} P_1(x)^5, 
\nonumber \\ 
 \ \ &=& -\frac{916731}{512}x^2 - \frac{33171}{128}x^4 - \frac{6453}{256}x^6 
   -\frac{823}{512}x^8 - \frac{319}{5120}x^{10} - \frac{7}{6144}x^{12},
 \nonumber \\ 
f_6(x) &=&P_6(x) -P_1(x) P_5(x) - P_2(x) P_4(x)+P_1(x)^2 P_4(x)
 -\frac{1}{2} P_3(x)^2 + 2 P_1(x) P_2(x) P_3(x) 
  \nonumber \\ 
 \ \ \ & & -P_1(x)^3 P_3(x)+\frac{1}{3} P_2(x)^3 - 
          \frac{3}{2} P_1(x)^2 P_2(x)^2 + 
      P_1(x)^4 P_2(x)  -\frac{1}{6} P_1(x)^6,\nonumber \\ 
 \ \ &=& \frac{65518401}{2048}x^2 + \frac{19425763}{4096}x^4 +
  \frac{752825}{1536}x^6+\frac{43783}{4096}x^8 + \frac{3481}{2048}x^{10}
    +  \frac{1255}{24576}x^{12} \nonumber \\ 
 \ \ & &  + \frac{3}{4096}x^{14}, 
\eeq 
 and so on.
$f_1(x)\sim f_3(x)$ coincide with the results in Ref. 10), where
explicit expressions of $f_n(x)$ are given  only for $n\leq 3$.
 It is interesting that the polynomials $f_n(x)$ are given in terms of 
 $P_n(x)$  appearing in the naive perturbative expansion in a closed 
form.
It should be instructive to remark here that $f_i(x)$'s 
($i=1, 2, 3, ...$) are the {\em cumulant}\cite{cumulant} of the sum 
$\sum_{n=0}^{\infty}\eps^nP_n(x)$ in the sense that 
\beq
\sum_{n=0}^{\infty}\eps^nP_n(x)\sim {\rm exp}
[\sum_{n=0}^{\infty}\eps ^nf_n(x)].
\eeq

In short, the RG method based on the construction of an envelope
 certainly  resumes the perturbation series of the 
 wave function and the resultant expression are given 
 in terms of  the cumulants of the naive perturbation series. 
 Conversely, our method provides a mechanical way
 for computing cumulants of given series.

\subsection{Reproducing the WKB result}
We now examine  how the
 WKB result Eq.(\ref{wkb}) can be reproduced from the perturbation series
 obtained above.
 Bender and Bettencourt found that if all terms beyond 
 $1/512\cdot \eps^4 x^{10}$ are neglected,
 the sum of the highest power terms in 
 $f_j(x)$ $(j\leq 4)$ is nicely rewritten as 
$-{x^2}/{4}(1+2\eps x^2+{17}/{12}\cdot\eps^2x^4+{5}/{12}\cdot\eps^3 x^6
 +{47}/{1152}\cdot\eps^4 x^8) ^{1/8}$,
which behaves for large $x$ as 
\beq
-\sqrt{\eps}\vert x\vert ^3/4(1152/47)^{1/8} \simeq
\sqrt{\eps}\vert x\vert ^3/5.96663,
\eeq
 in an excellent agreement with the WKB result.
How about the higher orders.
In the fifth order, the sum of the highest powers may be rewritten 
by neglecting all terms beyond $7\eps^5 x^{10}/1286$ as
$-{x^2}/{4}\cdot(1+ \eps x^2/4 -\eps^2 x^4/24 + \eps ^3 x^6/64
-\eps ^4 x^8/128  + 7\eps^5 x^{10}/1286)$
$\sim -{x^2}/{4}(1+{5}/{2}\cdot\eps x^2 +{115}/{48}\cdot\eps^2 x^4
 + {35}/{32}\cdot\eps^3 x^6+{15}/{64}\cdot\eps^4 x^8 + {4459}/{164608}\cdot
\eps^5 x^{10})^{1/10}$.
For large $x$, the coefficient of $-\sqrt{\eps}\vert x\vert^3$ is
$4(164608/4459)^{1/10}\simeq 5.73827$,
 which deviates from 6 more 
 than the fourth order result. Unfortunately, the sixth order becomes worse:
The sum of the highest powers is rewritten as
$ -{x^2}/{4}(1+3\eps x^2 +{29}/{8}\cdot\eps^2 x^4
 + {9}/{4}\cdot\eps^3 x^6+{577}/{768}\cdot\eps^4 x^8
 +{67621}/{493824}\cdot\eps^5 x^{10} +{1324349}/{35555328}\cdot
\eps^6x^{12})^{1/12}$,
which makes the coefficient of $-\sqrt{\eps}\vert x\vert^3$ for large $x$ 
\beq
4(35555328/1324349)^{1/12}\simeq 5.26181.
\eeq
This is actually plausible because the convergence radius of the
 perturbation series is zero; the cumulant series should be  at best  
 an asymptotic series. 

\section{Summary and Concluding Remarks}
In summary, we have successfully  applied the RG method\cite{cgo} as 
 formulated in Ref.~\citen{kuni1,kuni2} 
 to Schr\"{o}dinger equation of the quantum anharmonic oscillator:
 The naive perturbation series of the wave function are 
resummed by the RG equation. We have found the following:
Although the sum of the highest power in $f_n(x)$
 can be organized so that it becomes asymptotically  proportional
 to $\sqrt{\eps}\vert x\vert^3$ as was done  in \cite{bb,gm}, 
the coefficient of it  
 reaches the  closest value  to 6, the WKB result, in the fourth order,
 then  goes away monotonously from the closest value in the higher
 orders. We remark that the RG method as developed here can be also applied to
 the first excited state.

It should be stressed that the method presented here can improve
 perturbative wave functions   for all cases where naive perturbative 
solutions of Bender-Wu type  are given; 
namely, the RG method combined with the Bender-Wu
 perturbation method constitutes a new powerful method for improving
 perturbative wave functions. 
 It is interesting to extend the present method 
  and apply it to quantum  field theory, although it has been indicated that 
 the notion of envelopes is useful for improving the effective potential
 given in the loop expansion.\cite{kuni1}

The present RG method is an application of the perturbative RG equation.
 Can  nonperturbative RG equations be useful to construct global wave 
functions?  It should be. Indeed a variational 
 method called the delta-expansion method\cite{delta} which also 
 utilizes an RG-like equation but nonperturbatively has been 
 extended for obtaining wave functions.\cite{prl}\  
 The key ingredient of the extension is to construct an envelope of a set
 of  perturbative wave functions as in the RG method but 
{\em with a variational parameter}.
In this method, although the basic equation  can not be solved analytically but
 only numerically, 
  uniformly valid wave functions  with correct asymptotic behavior
 are obtained in the first-order 
 perturbation even for strong couplings and for excited states.
In the present method, the basic equations are solved analytically,
 and the asymptotic form of the  wave function is constructed explicitly,
 although a further resummation devised  in \cite{bb,gm} 
is needed for obtaining the asymptotic
 form. In this sense, the two methods are complementary to each other.
 It would be interesting if one can  combine the two methods.

\section*{Acknowledgement}
This work was partially supported by the Grants-in-Aid of the Japanese Ministry
 of Education, Science and Culture (No. 09640377).


\end{document}